# ESTUDIO COMPARATIVO ENTRE LAS DIFERENTES FUENTES DE ENERGÍA ELÉCTRICA EN COLOMBIA Y LA GENERACIÓN DE ELECTRICIDAD A PARTIR DE BIOMASA

Juan Mauricio García Torres
Fundación Universitaria Agraria de Colombia
Bogotá D.C

*Abstract*—. **This paper presents a comparison about different ways to generate electricity in Colombia (mainly the operation of each method) with respect to a technique made in Europe, focused in the develop of energy with biomass or solid waste, taking into account that Colombia is a country with an economy based in agriculture processes. The article presents the advantages and disadvantages in each method with variables like Generation, investment, employment and environment aspects, and with these information a proposal is presented like the best way to generate energy in the country.**

## I. INTRODUCCION

En el mundo actual, la energía eléctrica ha beneficiado a la humanidad mejorando su calidad de vida. Está presente en hogares, industrias, calles e interviene desde la producción de manufactura, hasta en los métodos de comunicación, en la forma de preparar alimentos y en el funcionamiento de diversos dispositivos que facilitan la realización de las acciones cotidianas. Sin embargo, algunos métodos que permiten generar la energía necesaria para obtener todos estos beneficios, contribuyen de manera negativa al medio ambiente, afectando gravemente la calidad de vida de la población; aumentando el número de muertes prematuras, atribuidas a enfermedades respiratorias y cardiovasculares, por lo que se debe optar por métodos de producción de energía más amigables con el medio ambiente (fedebiocombustibles, 2013).

En el desarrollo de este artículo se presenta un análisis de los diferentes métodos de energía eléctrica presentes en Colombia, resaltando aspectos de producción, funcionamiento, impacto social, ambiental, y se propone la implementación de una central eléctrica convencional de biomasa como fuente de energía eléctrica, que permita de una manera renovable, contribuir al sistema nacional de interconexión (SIN).

## II. SECTOR ELECTRICO EN COLOMBIA

Colombia es un país que posee una gran ventaja con respecto a la ubicación y al papel de la agricultura como principal fuente de ingresos económicos, permitiendo obtener energía eléctrica de diferentes medios (combustibles fósiles, parques eólicos, centrales hidroeléctricas y biomasa).

Según Fedesarrollo: "Colombia posee una ubicación privilegiada que le permite la explotación de recursos hídricos para la generación de electricidad. Desde los comienzos de la producción de electricidad en el país se aprovechó la abundante presencia de cuencas hídricas y el pronunciado relieve del país; condiciones ideales para el aprovechamiento de este recurso. Es debido a esta condición que la matriz eléctrica colombiana presenta una composición totalmente diferente a la matriz eléctrica mundial, en la cual predomina la generación a partir de combustibles fósiles. En el caso colombiano esta es dominada por la generación hídrica". [1].

La empresa Colombia Energía afirma lo siguiente: "Actualmente Colombia cuenta con más de 30 centrales hidroeléctricas, lo cual ha permitido consolidar al país como el quinto país más competitivo en generación energética, y estas plantas participan en el 63% de la energía que se suministra en todo el país" [2].

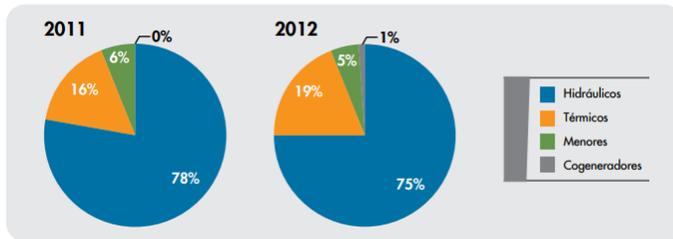

Fig. 1. Generación de electricidad en Colombia [3].

Teniendo en cuenta que solo hay 10 plantas de Cogeneración en el país, el aporte que realizó al Sistema de Interconexión Nacional (SIN) fue de 1% y que las plantas de cogeneración producen un 40% electricidad, significa que tuvo un buen desempeño en generación de energía eléctrica.

*A. Centrales hidroeléctricas*

Uno de los métodos más antiguos para la generación de energía eléctrica es a partir del movimiento del agua, sea a partir de caídas de ríos o en las mareas. Las centrales hidroeléctricas son aquellas en las que aprovechan la energía cinética del agua para transformarla en electricidad a partir de un generador.

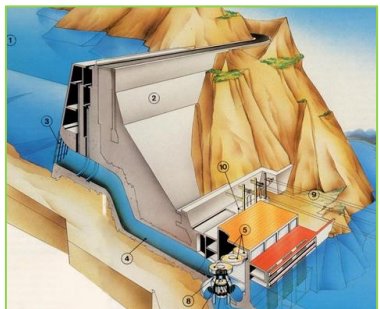

Fig. 2. Esquema de una central hidroeléctrica [4]

Según Endesa Educa [5], actualmente se considera que la energía eléctrica proveniente de centrales hidroeléctricas es limpia, pero, según una investigación de Endesa Educa, se afirma que las consecuencias generadas por la creación de centrales hidroeléctricas son las siguientes:

• Alteración del territorio
• Modificación del ciclo de vida de la fauna.
• Dificultad en la navegación fluvial y el transporte de materiales aguas abajo (nutrientes y sedimentos, como limos y arcillas).
• Disminuye el caudal de los ríos, modificando el nivel de las capas freáticas, la composición del agua embalsada y el microclima.

*B. Centrales termoeléctricas*

Las centrales termoeléctricas son aquellas en las que se obtiene energía eléctrica a partir de combustibles fosiles, tales como el carbón, el gas natural y el petróleo.

El funcionamiento de una central termoeléctrica se basa principalmente en la combustión de la materia prima en una caldera, la cual genera energía térmica que calienta unas toberas con agua, convirtiendo el agua en vapor y gracias a presión del fluido, permite el movimiento de las aspas de una turbina conectada a un generador, transformando la energía mecánica en energía eléctrica.

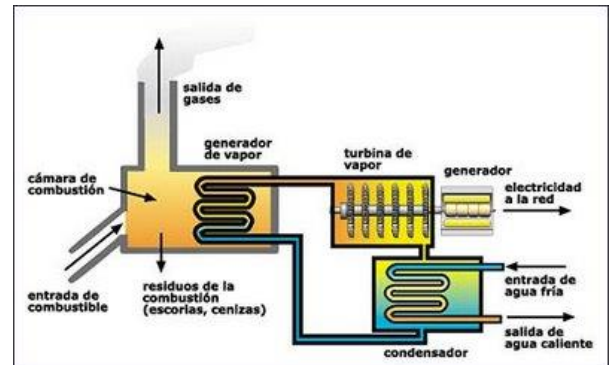

Fig. 3. Esquema de una central termoeléctrica. [6]

Según una investigación de Endesa Educa [7], entre las desventajas más importantes del uso de las termoeléctricas se resaltan las siguientes:

• Emisión de residuos a la atmósfera: Este tipo de residuos provienen de la combustión de los combustibles fósiles que utilizan las centrales térmicas convencionales para funcionar y producir electricidad. Esta combustión genera partículas que van a parar a la atmósfera, pudiendo perjudicar el entorno del planeta.

• Transferencia térmica: Algunas centrales térmicas (las denominadas de ciclo abierto) pueden provocar el calentamiento de las aguas del río o del mar. Este tipo de impactos en el medio se solucionan con la utilización de sistemas de refrigeración, cuya tarea principal es enfriar el agua a temperaturas parecidas a las normales para el medio ambiente y así evitar su calentamiento.

*C. Parques eólicos*

Un parque eólico es un conjunto de aerogeneradores capaces de aprovechar el viento para generar electricidad. Una de las desventajas presentes en este tipo de generación de energía eléctrica es la dependencia a las situaciones meteorológicas del sector.

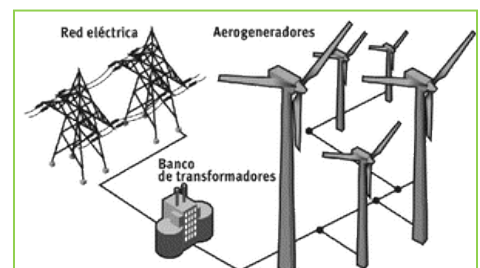

Fig. 4. Esquema de una central hidroeléctrica. [8]

Según una investigación presentada por Energy Spain en el año 2012 [9], las desventajas que presenta la energía eólica son las siguientes:

• Las hélices de los aerogeneradores pueden afectar las aves que se encuentren en el entorno
• Contaminación Acústica generada por el giro de las turbinas
• Dificultad en la zona de instalación de un parque eólico

*D. Planta de cogeneración de biomasa*

Las plantas de cogeneración son aquellas que convierten la materia prima (sea combustibles fósiles o biomasa) en energía eléctrica y energía térmica. Existen alrededor de 10 plantas de cogeneración de biomasa en Colombia y han aportado un 1% al Sistema de Interconexión Nacional (SIN).

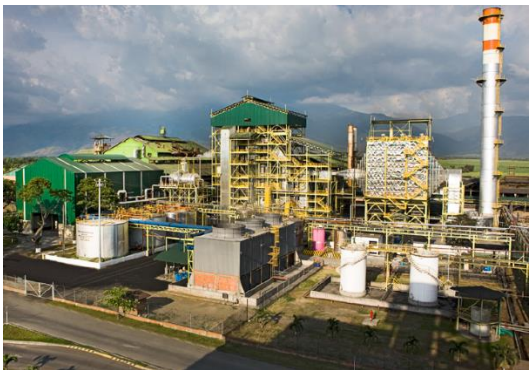
Fig. 5. Central de cogeneración. [10]

## III. BIOMASA

Aldoma afirma: "La biomasa es un material orgánico que proviene de un proceso biológico o industrial el cual se considera como desecho o residuo y se utiliza para generar energía de manera renovable" [11]

*A. Tipos de biomasa*

La biomasa se puede clasificar de la siguiente forma:

• **La biomasa natural:** Es aquella que se produce en la naturaleza sin intervención humana.
• **La biomasa residual:** Es el subproducto o residuo generado en las actividades agrícolas, silvícolas, ganaderas, residuos de la industria agroalimentaria y en la industria de transformación de la madera, así como residuos de depuradoras y del reciclado de aceites.
• **Los cultivos energéticos:** Son aquellos que están destinados a la producción de biocombustibles.

*B. Ventajas del uso de la biomasa*

Según una investigación hecha por Twenergy en el año 2012 [12], la biomasa presenta las siguientes ventajas:

• La transformación de un desecho en un recurso, realizando un aumento en el reciclaje y una disminución en la cantidad de residuos sólidos.
• La no contribución al cambio climático: su balance en emisiones de $CO_2$ es neutro. Al quemar la biomasa para obtener energía se libera $CO_2$ a la atmosfera, pero durante el crecimiento de la materia orgánica vegetal se absorbe el $CO_2$, permitiendo un balance entre el nivel de emisión y el nivel de aprovechamiento del gas por la naturaleza.
• Al realizar el proceso de combustión no provoca el fenómeno de la lluvia ácida.
• Con respecto al sector económico, el precio de la biomasa con respecto al del petróleo es mucho menor.

## IV. GENERACIÓN DE ENERGÍA ELÉCTRICA A PARTIR DE UNA PLANTA DE BIOMASA CONVENCIONAL

En Colombia no se ha implementado un proyecto relacionado con la construcción de una planta convencional de energía eléctrica de biomasa, ya que las empresas deciden apuntar a la Cogeneración como método de obtención de electricidad. Se evidencia la necesidad de implementar un sistema de generación de energía eléctrica a partir de la biomasa, con el fin de satisfacer las necesidades de la población. Esto generaría un impacto positivo con respecto al tratamiento de residuos sólidos al aprovechar mejor los desperdicios generados por procesos agroindustriales y a la vez, aumentaría la calidad de vida de las personas que habitan dicho sector.

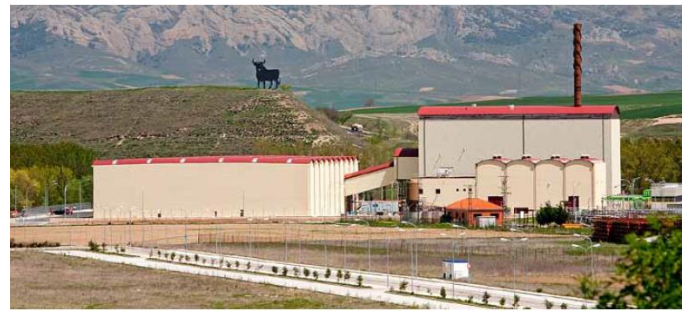
Fig. 6. Central de biomasa. [13]

Según TODOSBIOMASA.COM, en los lugares donde se encuentran implementadas estas centrales han generado resultados satisfactorios y se ha beneficiado a la sociedad a través de los residuos sólidos que se generan en sus procesos.

*A. Funcionamiento de una planta de biomasa*

Las plantas de biomasa funcionan de la siguiente forma:

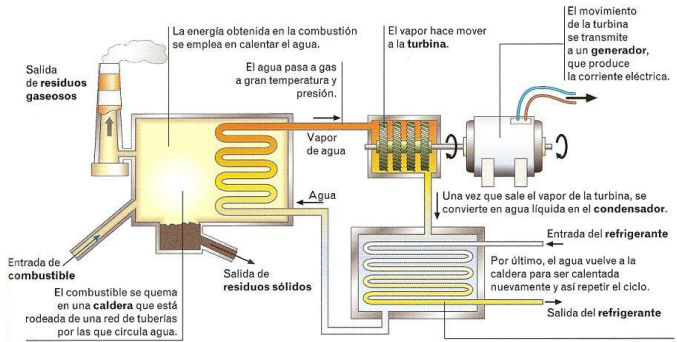

Fig 7. Funcionamiento de una central de biomasa [14]

En la figura anterior se observa el funcionamiento de una planta eléctrica de biomasa, allí la biomasa llega a la planta donde es separada según su tamaño. En la caldera la biomasa se quema controlando las variables presentes en el proceso de combustión (temperatura, flujo de aire, etc.) para mantener siempre las condiciones óptimas. Ese calor producido por la combustión de la biomasa convierte el agua que pasa por unas tuberías en vapor. Ese vapor de agua pasa por unas toberas que aumentan la presión del vapor. Ese flujo de vapor mueve una turbina que convierte la presión del fluido en energía mecánica. Un generador aprovecha esa energía para producir electricidad. La energía eléctrica creada por el generador se dirige a un transformador en el cual pasa a la red eléctrica convencional. El vapor se condensa y el agua una vez tratada vuelve a ser utilizada en el proceso. Los gases resultantes de la combustión de biomasa son filtrados para evitar en lo posible la contaminación de la atmosfera.

## V. Comparación entre fuentes de energía eléctrica

Se puede realizar una comparación entre los diferentes métodos de obtención de energía eléctrica teniendo en cuenta diferentes variables o puntos de vista, que permitirán seleccionar el tipo de método óptimo para implementarse en de Colombia. Se debe tener en cuenta que en la comparación se seleccionaron métodos con una capacidad de generación de electricidad.

Tabla I.
Variables de las diferentes fuentes de electricidad

|  | Central hidroeléctrica | Central termoeléctrica | Parque eólico | Planta de cogeneración | Planta convencional de biomasa |
|---|---|---|---|---|---|
| Capacidad de generación (MW) | 19 | 30 | 19.5 | 36 MW en total (26 para consumo interno, 10 de venta) | 16 |
| Generación Anual estimada (GWh/año) | 174 | 201 | 85 | 220 (70 entregado al SIN) | 128 |
| Inversión (Millones de $) | 127.000 | 45.000 | 55.600 | 160.000 | 120.000 |
| Cantidad de empleos generados | 300 | 60 | 38 | 400 | 100 |

Fuentes: [15], [16], [17] y [18]

### A. Capacidad de generación de energía eléctrica

A continuación se presenta un análisis comparativo sobre la capacidad de generación de energía eléctrica según los diferentes métodos como: centrales hidroeléctricas, centrales termoeléctricas, parques eólicos, plantas de cogeneración y plantas de biomasa.

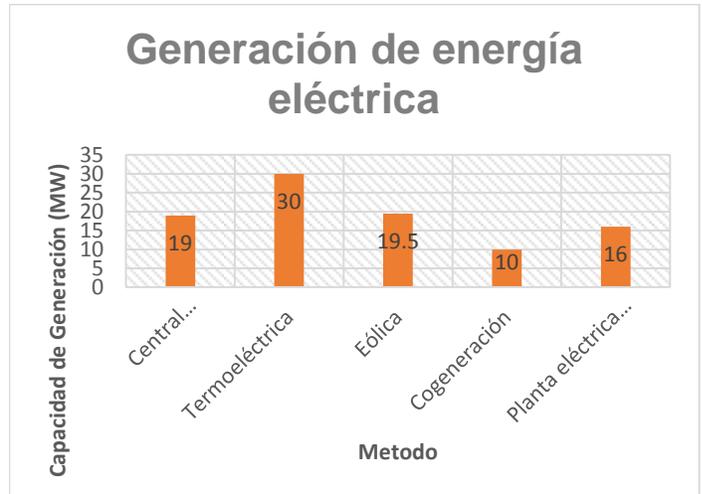

Fig 8. Comparación según la capacidad de generación de energía eléctrica.

Teniendo en cuenta la figura anterior y la generación promedio de energía de 23MW, se observa que la planta eléctrica de biomasa presenta una mayor capacidad de generación de energía con respecto a las plantas de cogeneración. En relación con los demás métodos de generación de electricidad, se presenta una menor capacidad de generación de energía, pero cabe resaltar que estas fuentes de energía presentan dificultades tales como la dependencia del entorno donde se encuentran ubicadas (las centrales hidroeléctricas dependen del recurso hídrico y los parques eólicos dependen de las condiciones meteorológicas del ambiente), los costos de la materia prima (el precio de los combustibles fósiles es mayor al precio de la biomasa y no son renovables).

### B. Generación anual estimada

Teniendo en cuenta la información relacionada a la generación anual estimada de energía, se puede realizar el siguiente histograma

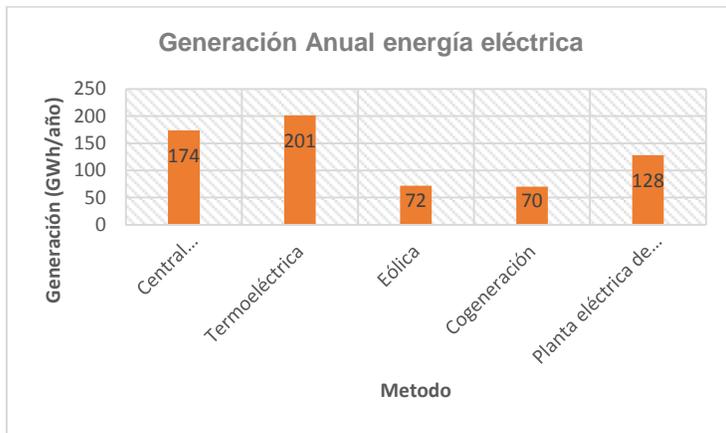

Fig 9. Comparación según la generación anual estimada de energía eléctrica.

Según la figura anterior y la generación promedio de energía de 23MW, se observa que la planta de generación de energía eléctrica a partir de biomasa posee una mayor capacidad de generación de energía anual con respecto a una planta de biomasa con sistema de cogeneración y a los parques eólicos. En este caso, se presentan variaciones con respecto a los parques eólicos, debido a que este método depende principalmente de los factores ambientales y meteorológicos, en los que puede variar la obtención de energía según la variación de la velocidad del viento en diferentes épocas

*C. Inversión*

Con respecto a la inversión que se debe realizar para el desarrollo de un proyecto en relación con la energía eléctrica, se puede comparar en el siguiente histograma.

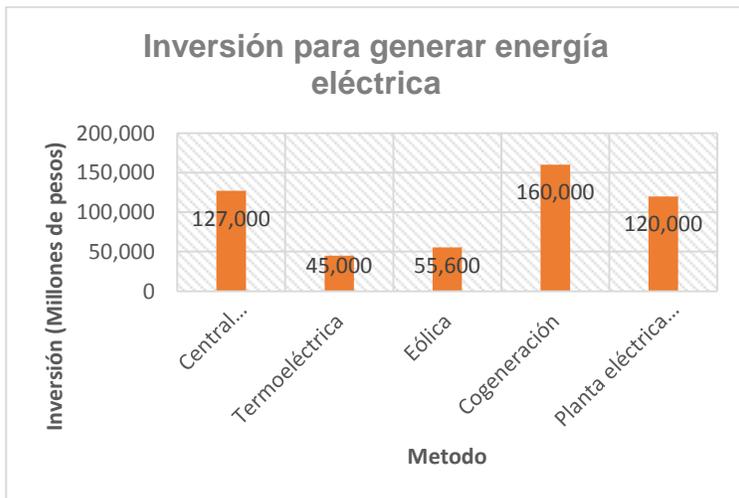

Fig 10. Inversión necesaria para la construcción de una central generadora de electricidad según diferentes métodos de obtención de energía.

Según la figura anterior y la generación promedio de energía de 23MW, las centrales termoeléctricas son los procesos que menor inversión necesitan, junto a los parques eólicos, pero como consecuencia, los gastos que se presentan con respecto a la obtención de materia prima en centrales termoeléctricas son elevados y los parques eólicos no poseen una generación anual elevada. Las plantas eléctricas de biomasa requieren menor inversión con respecto a las centrales hidroeléctricas y a las plantas de cogeneración.

*D. Empleo*

La siguiente grafica muestra la cantidad de empleo que se genera para el funcionamiento de las diferentes centrales de generación de energía eléctrica.

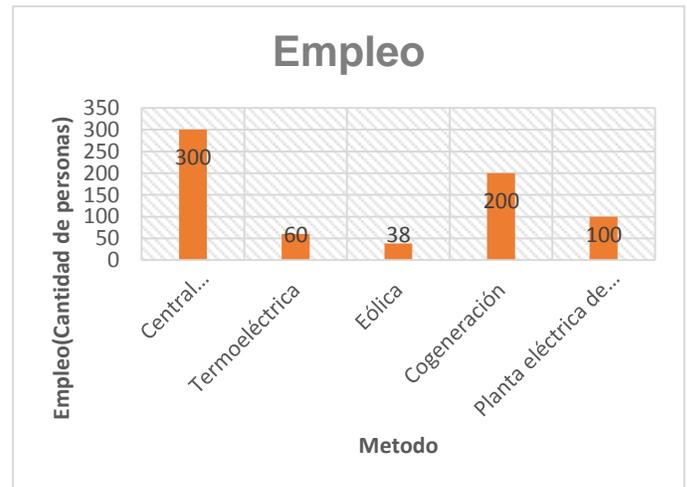

Fig 11. Mano de obra necesaria para las diferentes centrales de energía eléctrica.

Según la figura anterior y la generación promedio de energía de 23MW, se observa una descripción cuantitativa de la mano de obra requerida en el funcionamiento de diferentes métodos de generación de energía. Es evidente que en los parques eólicos el número de personal es menor en relación a los demás métodos de generación de electricidad. Esta afirmación puede ser interpretada desde dos puntos de vista:

Desde el punto de vista social, al construir un parque eólico se aumenta el índice de desempleo en esta actividad, afectando negativamente al sector económico donde se encuentra implementado el sistema eólico.

El mantenimiento de un parque eólico es costoso a causa de que no hay una supervisión constante del sistema.

*E. Aspectos ambientales*

El efecto invernadero es aquel fenómeno donde los gases que se encuentran en el planeta retienen el calor y no permite liberar energía al espacio. El exceso de gases de efecto invernadero es lo que provoca el calentamiento global.

El siguiente grafico presenta el porcentaje de los gases que producen el efecto invernadero

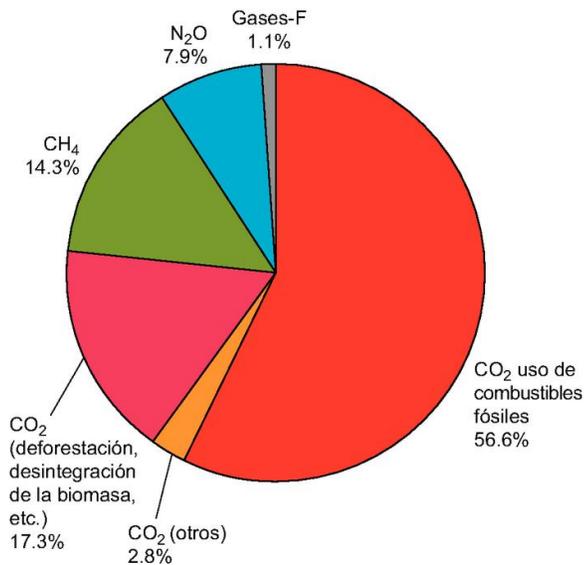

Fig 12. Participación de los gases en el efecto invernadero [19]

El Dióxido de Carbono ($CO_2$) es uno de los gases con mayor participación en el efecto invernadero. Según la compañía Inspiration [20], las consecuencias que se generan gracias al efecto invernadero son las siguientes:

- Deforestación
- Desertización
- Inundaciones
- Huracanes, tifones
- Sequía
- Fusión de los casquetes polares

## VI. RESULTADOS DE LA COMPARACIÓN

Teniendo en cuenta que el CO2 es el gas con mayor participación en el efecto invernadero, se debe resaltar que los procesos relacionados con biomasa generan dióxido de carbono de manera neutral. La empresa Twenergy afirma lo siguiente: "la cantidad de CO2 que las plantas absorben es la misma que se emite en su combustión y por lo tanto, su uso energético no contribuye al aumento del porcentaje del gas en la atmosfera [12].

Teniendo en cuenta todos los aspectos anteriores se puede seleccionar como opción viable la planta de energía eléctrica convencional a partir de biomasa por las siguientes razones:

• Las plantas de energía eléctricas convencionales a partir de biomasa no dependen necesariamente del entorno para su ejecución, por lo que se pueden construir lejos de un rio (a diferencia de una central hidroeléctrica) y no depende de condiciones meteorológicas para su funcionamiento (parques eólicos).

• Las plantas de energía eléctrica convencionales a partir de biomasa presentan una alta estimación anual en generación de energía (128GWh/año) a comparación con los parques eólicos (72GWh/año) y las plantas de cogeneración (70GWh/año).

• La biomasa no contribuye al aumento del CO2 en la atmosfera, por lo que no contribuye a la producción del principal gas relacionado con el efecto invernadero.

• A pesar de que las plantas de energía eléctrica convencionales a partir de biomasa presentan una inversión alta con respecto a los parques eólicos, este último método tiene una estimación anual de 72GWh/año y las plantas de biomasa convencionales tienen una estimación anual de 128Gwh/año, quiere decir que los parques eólicos deben realizar una inversión del más del 100% para alcanzar la estimación anual cercana a la de las plantas de biomasa, por lo tanto se puede justificar el precio de inversión de una planta eléctrica de biomasa.

## VII. CONCLUSIONES

Todos los métodos para obtención de energía eléctrica contaminan, expulsando gases o partículas al ecosistema, pero el impacto ambiental que genera cada fuente de energía es lo que se debe tener en cuenta para minimizar los perjuicios que han sido ocasionados por los procesos tradicionales de obtención de electricidad y empezar a implementar las energías alternativas que permitan obtener la energía eléctrica necesaria para las personas a un precio menor para el ambiente.

La biomasa es una de las fuentes más importantes de generación de electricidad en el mundo, aprovechando los residuos sólidos para obtener energía eléctrica. Colombia es un país en el que posee un gran potencial con respecto al aprovechamiento de residuos sólidos, ya que la base de su economía es la agroindustria, biomasa capaz de ser aprovechada para beneficiar a ciertas zonas del país que no poseen electricidad.

La manera en la que funcionan las plantas de biomasa y las termoeléctricas es similar, sólo cambia la fuente de energía. En este sentido, los residuos que se obtienen de la biomasa son utilizados como fertilizantes para abono y no generan CO2, siendo una opción viable en temas medioambientales.

## VIII. REFERENCIAS